\documentclass[journal]{IEEEtran}

\usepackage[utf8]{inputenc}
\usepackage[english]{babel}
\usepackage{fancyhdr}
\usepackage{graphicx}
\usepackage{subfigure}
\usepackage{amsmath}
\usepackage{epstopdf}
\usepackage{booktabs}
\usepackage[english]{babel}
\usepackage{amssymb}
\usepackage{tipa}
\usepackage{lpic}
\usepackage{stfloats}
\usepackage{algorithmic}
\usepackage{url}
\usepackage{verbatim}
\usepackage{empheq}
\usepackage{cases}
\usepackage{multirow}
\usepackage{booktabs}
\usepackage{threeparttable}
\usepackage[ruled,vlined]{algorithm2e}

\usepackage{flushend}
\usepackage{setspace}

\usepackage[noadjust]{cite}
\usepackage{enumerate}
\usepackage{amsthm}

\ifCLASSINFOpdf

\else

\fi

\begin{document}

\title{DDPG-based Resource Management for MEC/UAV-Assisted Vehicular Networks}

\author{\IEEEauthorblockN{Haixia Peng,~\IEEEmembership{Student Member,~IEEE}, Xuemin (Sherman) Shen,~\IEEEmembership{Fellow,~IEEE}\\
\IEEEauthorblockA{Department of Electrical and Computer Engineering, University of Waterloo, Waterloo, ON, Canada, N2L 3G1}\\
Email: \{h27peng, sshen\} @uwaterloo.ca}\vspace{-2mm}
}


\maketitle
\IEEEpeerreviewmaketitle

\begin{abstract}
In this paper, we investigate joint vehicle association and multi-dimensional resource management in a vehicular network assisted by multi-access edge computing (MEC) and unmanned aerial vehicle (UAV). To efficiently manage the available spectrum, computing, and caching resources for the MEC-mounted base station and UAVs, a resource optimization problem is formulated and carried out at a central controller. Considering the overlong solving time of the formulated problem and the sensitive delay requirements of vehicular applications, we transform the optimization problem using reinforcement learning and then design a deep deterministic policy gradient (DDPG)-based solution. Through training the DDPG-based resource management model offline, optimal vehicle association and resource allocation decisions can be obtained rapidly. Simulation results demonstrate that the DDPG-based resource management scheme can converge within $200$ episodes and achieve higher delay/quality-of-service satisfaction ratios than the random scheme.
\end{abstract}


\maketitle

\IEEEpeerreviewmaketitle

\section{Introduction}
\label{sec:Intro}

Vehicular networks, referred to the wireless networks for vehicles to everything communications, are attracting more and more attention from academia and industries and are significantly improving intelligent transportation services \cite{gurugopinath2019cache, liang2019spectrum, ye2019deep, liang2019deep}. Via vehicular networks, road safety and traffic efficiency are enhanced and an increasing number of vehicular applications and data services are enabled \cite{peng2019spectrum}. However, due to limited spectrum resources \cite{zhang2014dynamic} and on-board computing/caching resources, propelling vehicular networks to support the emerged applications and services, especially the one with tasks requiring sensitive-delay and multiple dimensions of resources, still faces a host of challenges \cite{tayyaba20205g, zhang2013cooperative}. To address those challenges while controlling the costs on resource deployment in an acceptable range, multi-access edge computing (MEC) and unmanned aerial vehicle (UAV) technologies have been applied to vehicular networks \cite{Peng2019SDN, zhao2019computation, chen2019uav, zhang2018energy, LEAD2020feng}. 

By shifting computing and caching resources to edge nodes or base stations to enable preset MEC servers in vehicular networks, vehicles can choose to offload some tasks to the MEC servers via different access technologies. As data transmission between the MEC server and the core network is avoided, compared to tasks offloaded to the cloud computing server, a shorter response delay is provided to satisfy the sensitive delay requirement of each task offloaded to the MEC server \cite{Peng2019SDN, ning2019mobile}. Moreover, considering the time-varying demanding on resources in vehicular networks, especially with bursty traffic caused by some events or social activities, insufficient resource issues remain on the MEC-mounted base station. Increasing the amount of resources installed at each base station physically would cause waste most of the time. Thus, in addition to MEC-mounted base stations, removable and flexible MEC servers can be enabled by mounting MEC servers in UAVs to assist the bursty resource-demanding in vehicular networks \cite{chen2019uav, yang2019energy}. 

To benefit the vehicular networks from the MEC-mounted base stations and UAVs, plenty of existing works have been proposed. For example, an air-ground integrated network has been proposed in \cite{cheng2018air} to deploy and schedule the MEC-mounted UAVs to support vehicular applications. To dynamically manage the available spectrum, computing, and/or caching resources for the MEC-mounted base stations or UAVs in vehicular networks, existing works have studied the computing task offloading and/or resource management schemes with or without considering the heterogeneous delay-requirements of vehicular applications \cite{peng2019spectrum, peng2020deep, ning2019mobile}. However, most of the existing works have targeted at the vehicular scenarios only with the MEC-mounted base stations or UAVs. How to simultaneously manage the available multi-dimensional resources for both MEC-mounted base stations and UAVs to support vehicular applications with heterogeneous and sensitive delay requirements still needs efforts.

In this paper, we investigate joint vehicle association and resource management in a vehicular network with MEC-mounted macro eNodeB (MeNB) and UAVs. To centrally manage the available multi-dimensional resources for the MEC-mounted MeNB and UAVs, assume a controller is installed at the MeNB and in charge of making vehicle association and resource allocation decisions for the whole network. Specifically, to maximize the number of tasks offloaded to the MEC servers while satisfying their quality-of-service (QoS) requirements, we formulate an optimization problem and execute it at the controller. Since the formulated problem is non-convex and high in compute complexity, we transform the formulated problem using reinforcement learning (RL) by taking the controller as an agent and design a deep deterministic policy gradient (DDPG)-based solution. Through offline training, optimal vehicle association and resource allocation decisions then can be made timely by the controller to satisfy the offloaded tasks' QoS requirements. 

The rest of this paper is organized as follows. The system model, including the MEC/UAV-assisted vehicular network model and the multi-dimensional resource management model, is described in Section \ref{sec:System_model}, followed with the formulated resource optimization problem. In Section \ref{sec:SADDPG_Mana}, we transform the formulated problem using RL and solve it with a DDPG-based algorithm. Extensive simulation results are provided in Section \ref{sec:Simulation_Res} to demonstrate the performance of the DDPG-based resource management scheme. Finally, we conclude this work in Section \ref{sec:Conclu}.

\section{System Model}
\label{sec:System_model}

In this section, we illustrate an MEC/UAV-assisted vehicular network model and a multi-dimensional resource management model, and then formulate a resource optimization problem.  

\subsection{MEC/UAV-Assisted Vehicular Network}
\label{subsec:vehi_network}

\begin{figure}[htbp]
\centering
\includegraphics[scale=0.32]{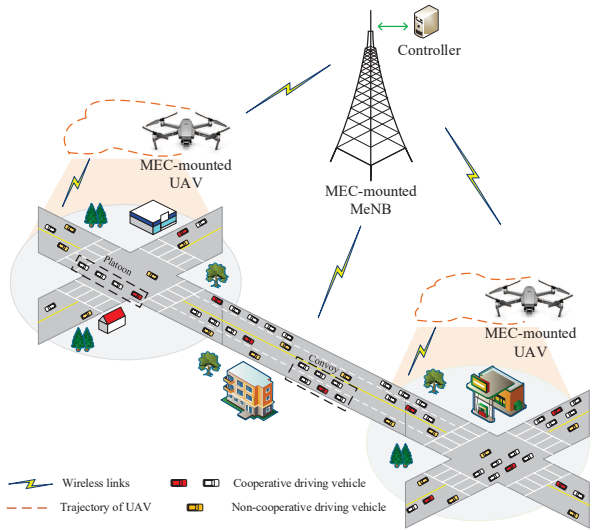}
\caption{An illustrative structure of the MEC/UAV-assisted vehicular network.}
\label{fig:systemTo}
\end{figure}

Consider a vehicular network with MEC-mounted MeNBs and UAVs to cooperatively support delay-sensitive and diverse resource-demanding vehicular applications. As illustrated in Fig. \ref{fig:systemTo}, an MeNB is deployed on one side of a straight two-way road, with two UAVs fly at a fixed speed above the road and on each side of the MeNB. The amounts of available spectrum, computing, and caching resources for the MeNB and UAV $j$ are denoted by $\{C^{sp}_m, C^{co}_m, C^{ca}_m\}$ and $\{C^{sp}_u, C^{co}_j, C^{ca}_j\}$ for $j = 1, 2$, respectively. Vehicles on the considered road segment randomly generate different computing tasks and send task offloading requests to the MeNB or UAV as needed. The task offloading request sent by vehicle $i$ at time slot $t$ is denoted by $\{c^{co}_i(t), c^{da}_i(t), c^{de}_i(t)\}$, where $c^{co}_i(t)$, $c^{da}_i(t)$, and $c^{de}_i(t)$ are the amount of computing resources required by, the data size of, and the maximum delay tolerated by the task, respectively. 


To efficiently manage the multi-dimensional resources of the MEC-mounted MeNB and UAVs among the received task offloading requests, a controller is enabled at the MeNB since the considered road segment is under the coverage area of the MeNB. The procedures of resource management then can be summarized as follows,
\begin{enumerate}
\item The controller receives driving information about and task offloading requests from vehicles under the coverage area of the MeNB;
\item According to the received information, a vehicle association and resource allocation decision is made by the controller to decide the association pattern\footnote{As the task division technology is not considered here, each vehicle is assumed to associate with and offload its computing task to the MeNB or one of the UAVs at each time slot.} for each vehicle and pre-allocate the available resources among the received task offloading requests;
\item The controller sends the computing and caching resource allocation results to each UAV and sends the spectrum allocation result and association pattern to each vehicle;
\item Each vehicle offloads its computing task to the MeNB or UAV over the allocated spectrum resources;
\item The MeNB and UAVs cache and process the received computing tasks and then return the processing results to the corresponding vehicles.  
\end{enumerate}

\subsection{Resource Management Model}
\label{subsec:res_model}

As the resource demand from offloaded tasks changes with time due to the high vehicle mobility and vehicle's heterogeneous computing tasks, the controller has to dynamically manage the available resources for the MeNB and UAVs. Denote the set/number of vehicles on the considered road segment at time slot $t$ as $\mathcal{N}(t)/N(t)$. $b_i$ is the association variable between vehicle $i\in \mathcal{N}(t)$ and the MeNB or the two UAVs, where $b_i = 1$ if vehicle $i$ associates with the MeNB and $b_i = 0$ otherwise. According to the location information about the MeNB, the two UAVs, and each vehicle, the controller can distinguish vehicles under different MEC servers. As each vehicle can only associate with one of the MEC servers, only vehicles under the coverage area of the MEC-mounted UAV would have $b_i = 0$. 

\textbf{Spectrum resources:} The controller allocates the $C^{sp}_m$ and $C^{sp}_u$ spectrum resources among vehicles associating with the MeNB and with the two UAVs, respectively. As the trajectory of each UAV is pre-designed, by controlling the distance between the two UAVs, spectrum reusing is adopted between uplink transmissions to the two UAVs with acceptable interference. Assume the transmit power of each vehicle for offloading computing tasks to the MeNB or UAV is fixed at $P$. Then the uplink transmission rates achieved by the MeNB and UAV $j$ from vehicle $i$ can be described as,
\begin{align}
        \label{Utility_MeNB}
         e_{m,i}(t) = C^{sp}_m f^{sp}_{m,i}(t){\rm log}_2 (1+ \frac{P G_{m,i}(t)}{\sigma^2}),
\end{align}
and 
\begin{align}
        \label{Utility_uav}
         e_{j,i}(t) = C^{sp}_u f^{sp}_{j,i}(t){\rm log}_2 (1 + \frac{P G_{j,i}(t)}{P G_{v,i}(t) + \sigma^2}),
\end{align}
respectively, where $v \in \{1,2\}/j$. $G_{m,i}(t)$ (or $G_{j,i}(t)$) is the gain of uplink channels from vehicle $i$ to the MeNB (or UAV $j$) at time slot $t$. The fractions of spectrum resources allocated to vehicle $i$ from the MeNB and UAV $j$ are denoted by $f^{sp}_{m,i}(t)$ and $f^{sp}_{j,i}(t)$, respectively.    

\textbf{Caching resources:} Vehicle $i$ sends its computing task with transmission rate $e_{m,i}(t)$ (or $e_{j,i}(t)$) to the MeNB (or UAV $j$). The data of each task has to be cached before processing the task. Let $f^{ca}_{m,i}(t)$ and $f^{ca}_{j,i}(t)$ be the fractions of caching resources allocated to vehicle $i$'s task from the MeNB and UAV $j$, respectively. Then vehicle $i$'s task can be successfully completed by the MEC server only if $C^{ca}_m f^{ca}_{m,i}(t) \geq c^{da}_i$ or $C^{ca}_u f^{ca}_{j,i}(t) \geq c^{da}_i$.    

\textbf{Computing resources:} Denote $f^{co}_{m,i}(t)$ and $f^{co}_{j,i}(t)$ as the fractions of computing resources allocated to vehicle $i$ from the MeNB and UAV $j \in \{1,2\}$, respectively. As the sizes of the offloading request and processing result are relatively small \cite{zhou2018computation}, we ignore the time cost on collecting requests from and sending the processing result back to each vehicle \cite{zhang2018energy}. For vehicle $i \in \mathcal{N}(t)$, the total time duration from generating the task to receiving the processing result then is,
\begin{align}
\begin{split}
\label{T_i}
T_i(t) = & \sum_{j\in\{1,2\}}(1-b_i)(\frac{c^{da}_i(t)}{e_{j,i}(t)} + \frac{c^{co}_i(t)}{C^{co}_u f_{j,i}^{co}(t)})\\ 
& + b_i(\frac{c^{da}_i(t)}{e_{m,i}(t)} + \frac{c^{co}_i(t)}{C^{co}_m f_{m,i}^{co}(t)}). 
\end{split}
\end{align}

\subsection{Problem Formulation}
\label{subsec:res_model}

Due to the heterogeneous computing tasks and high vehicle mobility, demanding on the multi-dimensional resources from offloaded tasks is time-varying. To dynamically manage the total available resources for the MeNB and the two UAVs to satisfy time-vary resource demand, an optimization problem is formulated. From both of the service provider and vehicle user's perspectives, it is critical to efficiently manage the finite resources to satisfy the resources demanding from as many offloaded tasks as possible. Thus, we formulate the optimization problem to maximize the number of tasks successfully completed by the MEC servers with given amounts of available resources. As a result, the problem is given as follows,
\begin{equation} \nonumber
\begin{split}
\label{problem_menb}
\max_{\substack{\textbf{b}(t), \textbf{f}^{sp}(t), \\ \textbf{f}^{co}(t), \textbf{f}^{ca}(t)}}   & \sum_{j\in\{1,2\}}\sum_{i\in\mathcal{N}(t)} \{ b_i(t) H[c^{de}_i(t)-T_i(t)] H[f_{m,i}^{ca}(t) \\ & C^{ca}_m -  c^{da}_i(t)] + (1-b_i(t))H[c^{de}_i(t)-T_i(t)] \\ & H[f_{j,i}^{ca}(t) C^{ca}_u - c^{da}_i(t)] \}  
\end{split}
\end{equation}
\begin{subnumcases}
{\rm s.t.}  
(\ref{Utility_MeNB}), (\ref{Utility_uav}), (\ref{T_i}) \label{Menb_Cons_1} \\
f_{m,i}^{sp}(t), f_{m,i}^{co}(t), f_{m,i}^{ca}(t) \in [0,1] \label{Menb_Cons_3} \\
f_{j,i}^{sp}(t), f_{j,i}^{co}(t), f_{j,i}^{ca}(t) \in [0,1], \quad j\in\{1,2\} \label{Menb_Cons_4} \\
b_i(t) \in \{0, 1\} \label{Menb_Cons_2} \\
\sum_{i\in\mathcal{N}(t)}b_i(t) f_{m,i}^{sp}(t) = 1  \label{Menb_Cons_5} \\
\sum_{i\in\mathcal{N}(t)}b_i(t) f_{m,i}^{co}(t) = 1  \label{Menb_Cons_6} \\
\sum_{i\in\mathcal{N}(t)}b_i(t) f_{m,i}^{ca}(t) = 1  \label{Menb_Cons_7} \\
\sum_{i\in\mathcal{N}(t)}(1-b_i(t)) f_{j,i}^{sp}(t) = 1, \quad j\in\{1,2\}  \label{Menb_Cons_8} \\
\sum_{i\in\mathcal{N}(t)}(1-b_i(t)) f_{j,i}^{co}(t) = 1, \quad j\in\{1,2\}  \label{Menb_Cons_9} \\
\sum_{i\in\mathcal{N}(t)}(1-b_i(t)) f_{j,i}^{ca}(t) = 1, \quad j\in\{1,2\},  \label{Menb_Cons_10}
\end{subnumcases}
where $\textbf{b}(t)$ is the association pattern matrix for vehicles in $\mathcal{N}(t)$. $\textbf{f}^{sp}(t)$, $\textbf{f}^{co}(t)$, and $\textbf{f}^{ca}(t)$ denote the allocation matrices of available spectrum, computing, and caching resources for the MeNB and UAVs, respectively. $H(\cdot)$ denotes the step function, which is $1$ if the variable is $\geq 0$, and $0$ otherwise. Then for vehicle $i$ with enough allocated caching resources for its computing task and satisfied task's delay requirement, we have $H[c^{de}_i(t)-T_i(t)] H[f_{m,i}^{ca}(t)C^{ca}_m -  c^{da}_i(t)]=1$ or $H[c^{de}_i(t)-T_i(t)]H[f_{j,i}^{ca}(t) C^{ca}_u - c^{da}_i(t)]=1$. The constraints on available resources for the MeNB and the two UAVs are given by (\ref{Menb_Cons_5})-(\ref{Menb_Cons_10}).

\section{DDPG-based Resource Management Scheme}
\label{sec:SADDPG_Mana}

From equation (\ref{T_i}) and the objective function in the optimization problem, the allocation of spectrum and computing resources is coupled with each other. Due to integer variable $b_i$ and the step function \cite{peng2020deep}, the formulated problem is non-convex. Also, the computational complexity of the problem increases with the number of vehicles under the coverage area of the MeNB. It is difficult to solve the formulated problem with the traditional optimization methods to satisfy tasks' strict delay requirements. Thus, to timely obtain a vehicle association and resource allocation decision, we transform the formulated problem using RL and then design a DDPG-based solution in this section. 

\subsection{Problem Transformation}
\label{subsec:Problem_Trans}

We transform the above optimization problem according to the main idea of RL \cite{liang2019deep, peng2020deep, shen2020ai}. Specifically, the controller implements the agent's role, and everything beyond the controller is regarded as the environment. Denote $\mathcal{S}$ as the environment state space. For each state $s \in \mathcal{S}$, the agent chooses association and resource allocation action $a$ from the action space $\mathcal{A}$ according to the current policy $\pi$, where $\pi:$ $\mathcal{S}$ $\mapsto$ $\mathcal{A}$. Then by executing $a$ in the environment, a reward $r$ will be returned back to the agent for guiding the policy updating until an optimal policy is obtained. Let $(x_i, y_i)$ and $(x^{\prime}_j, y^{\prime}_j, z^{\prime}_j)$ be the positions of vehicle $i \in \mathcal{N}(t)$ and UAV $j \in \{1, 2\}$, respectively. Then at time slot $t$, we can express the environment state as
\begin{align}
\begin{split}
        \label{env}
         s(t) = & \{c^{co}_1(t), c^{co}_2(t), \dots, c^{co}_{N(t)}(t), c^{da}_1(t), c^{da}_2(t), \dots, \\
         & c^{da}_{N(t)}(t), c^{de}_1(t), c^{de}_2(t), \dots, c^{de}_{N(t)}(t), x_1(t), x_2(t), \\
         & \dots, x_{N(t)}(t), y_1(t), y_2(t), \dots, y_{N(t)}(t), x^{\prime}_1(t), \\
         & x^{\prime}_2(t), y^{\prime}_1(t), y^{\prime}_2(t), z^{\prime}_1(t), z^{\prime}_2(t) \}.
\end{split}
\end{align}
Let $N_m^{\prime}$ and $N_j^{\prime}$ ($j \in \{1, 2\}$) be the numbers of vehicles associated to the MeNB and UAV $j$, respectively. Then the action selected by the agent at time slot $t$ is given by
\begin{align}   
\begin{split}
        \label{a}
         a(t) = & \{b_1(t), b_2(t), \dots,  b_{N(t)}(t), f_{m,1}^{sp}(t), f_{m,2}^{sp}(t), \\ & \dots, f_{m, N_m^{\prime}(t)}(t), f_{m,1}^{co}(t), f_{m,2}^{co}(t), \dots, \\ &  f_{m, N_m^{\prime}(t)}^{co}(t), f_{m,1}^{ca}(t), f_{m,2}^{ca}(t), \dots,   f_{m, N_m^{\prime}(t)}^{ca}(t), \\ & f_{j,1}^{sp}(t), f_{j,2}^{sp}(t), \dots, f_{j, N_j^{\prime}(t)}^{sp}(t), f_{j,1}^{co}(t), f_{j,2}^{co}(t), \\ & \dots,   f_{j, N_j^{\prime}(t)}^{co}(t), f_{j,1}^{ca}(t), f_{j,2}^{ca}(t), \dots,   f_{j, N_j^{\prime}(t)}^{ca}(t)\}.
\end{split}
\end{align}

As the agent updates policy $\pi$ according to the received reward $r$, to obtain an optimal policy that would achieve the objective function in the original problem, two reward elements are defined as follows,
\begin{align}
\label{r_de}
        r_i^{de}(t+1) =  {\rm log}_2 (\frac{c^{de}_i(t)}{T_i(t)} + 0.01)
\end{align}
\begin{align}
\label{r_ca}
r_i^{ca}(t+1) = 
\begin{cases}
{\rm log}_2 (\frac{ f_{m,i}^{ca}(t)C_m^{ca}}{c_i^{ca}(t)} + 0.01), &\mbox{if $b_i(t) = 1$}, \\
{\rm log}_2 (\frac{ f_{1/2,i}^{ca}(t)C_u^{ca}}{c_i^{ca}(t)} + 0.01), &\mbox{if $b_i(t) = 0$}, \\
\end{cases}
\end{align}
where, $r_i^{de}(t+1)$ and $r_i^{ca}(t+1)$ are the rewards achieved by vehicle $i$ from satisfied delay requirement and from allocated caching resources by action $a(t)$, respectively. Only if the delay requirement is satisfied and enough caching resources are allocated, positive rewards can be achieved by vehicle $i$, otherwise $r_i^{de}(t+1)$ and $r_i^{ca}(t+1)$ are negative. By using the logarithmic function and adding a small value $0.01$ in (\ref{r_de}) and (\ref{r_ca}), we can avoid sharp fluctuation on the rewards and therefore improving the convergence performance.     

\subsection{DDPG-based Solution}
\label{subsec:DDPG_Solu}

\begin{figure}[htbp]
\centering
\includegraphics[scale=0.36]{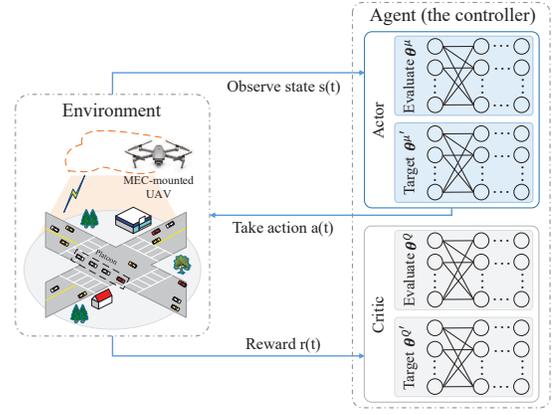}
\caption{The DDPG framework for the MEC/UAV-assisted vehicular network.}
\label{fig:SADDPG}
\end{figure}

The constraints of the original problem and equation (\ref{a}) indicate that the action space, $\mathcal{A}$, is continuous and discretizing $\mathcal{A}$ is infeasible due to the large size of each action. That is, the RL methods targeted at discrete state and action spaces, such as DQN, are inapplicable here. Thus, DDPG, an RL method that combines the advantages of policy gradient and DQN, is adopted to solve the transformed problem. As illustrated in Fig. \ref{fig:SADDPG}, the agent under DDPG is composed of an actor and a critic, both implemented by two deep neural networks (DNNs), i.e., a target network and an evaluation network. For an input environment state, the actor makes an action decision and the critic uses a Q function to value each pair of state-action. The standard Q-value function is given by
\begin{align}
\label{Q}
       Q(s,a) = \mathbb{E}[\sum_{\tau=0}^{\infty}\gamma^{\tau}r(t+\tau)|\pi, s=s(t), a=a(t)],
\end{align}
where $r(t)$ denotes the immediate reward returned to the agent at time slot $t$, which is defined as the average reward over vehicles in $\mathcal{N}(t)$, i.e., $r(t) = \frac{1}{N(t)}\sum_{i\in \mathcal{N}(t)}(r_i^{de}(t)+r_i^{ca}(t))$. And $\gamma$ is the discount factor on $r(t)$.


There are two stages of the DDPG-based solution, training and inferring, where the training stage is performed offline. As the correlation among transitions used in the training stage would reduce the convergence rate, experience replay technology is adopted in DDPG. Namely, saving $M_t$ transitions in a replay memory buffer first and then randomly selecting a mini-batch of transitions from the buffer to train the DDPG model, i.e., to update the actor and critic's parameters until converge. Parameters of the evaluation networks of the actor and the critic are updated according to policy gradient and loss function \cite{lillicrap2015continuous}, respectively. Specifically, the parameter matrix of actor's evaluation network, $\boldsymbol{\theta}^{\mu}$, is updated in the direction of $\nabla _{\boldsymbol{\theta}^{\mu}} J(\boldsymbol{\theta}^{\mu})$, where $\nabla _{\boldsymbol{\theta}^{\mu}}$ denotes the derivative w.r.t. $\boldsymbol{\theta}^{\mu}$ and $J(\boldsymbol{\theta}^{\mu})$ is the policy objective function. And the critic adjusts its evaluation network's parameters, $\boldsymbol{\theta}^{Q}$, in the direction of $\nabla _{\boldsymbol{\theta}^{Q}} L(\boldsymbol{\theta}^{Q})$ to minimize the loss, $L(\boldsymbol{\theta}^{Q}) = \mathbb{E}[(Q(s, a)|_t - (r(t) + \gamma Q^{\prime}(s, a)|_{t+1}))^2]$, where $Q^{\prime}(\cdot)$ is the Q-function of the critic's target network. With in-time updated parameters of the actor's and critic's evaluation networks, $\boldsymbol{\theta}^{\mu}$ and $\boldsymbol{\theta}^{Q}$, the agent then softly updates the parameters of the two target networks by
\begin{align}
        \label{soft_ac}
        \boldsymbol{\theta}^{\mu^{\prime}} = \kappa^a \boldsymbol{\theta}^{\mu} + (1 - \kappa^a)\boldsymbol{\theta}^{\mu^{\prime}}
\end{align}
and
\begin{align}
        \label{soft_cri}
        \boldsymbol{\theta}^{Q^{\prime}} =  \kappa^c \boldsymbol{\theta}^{Q} + (1 - \kappa^c)\boldsymbol{\theta}^{Q^{\prime}}
\end{align}
with $\kappa^a \ll 1$ and $\kappa^c \ll 1$, respectively.

Assume there are $M_s$ steps in one episode, and we use the total rewards achieved per episode, $r_{ep}$, to measure the change of rewards during the training stage. Then the DDPG-based solution can be summarized in Algorithm \ref{DDPG_Algorithm}. 

\begin{algorithm}
\SetAlgorithmName{Algorithm}{List of Models}
\DontPrintSemicolon
\tcc{Initialization}
{Initialize the size of the replay memory buffer and the parameter matrices of the two DNNs in both actor and critic.\\}
\tcc{Parameter updating}
 \ForEach{episode}{
Set initial state $s_0$ and let $r_{ep} = 0$. \\
 \ForEach{step $t$}{
Choose action $a(t)$ by the actor's evaluation network, i.e., $a(t) = \mu(s(t))$; \\
Obtain reward $r(t)$ and the subsequent state $s^{\prime}$; \\
 \uIf{the number of transitions $<$ $M_t$}{
Save transition $\{s(t), a(t), r(t), s^{\prime}\}$ in the replay buffer;
 }
 \Else{
Replace the first saved transition in the buffer with $\{s(t), a(t), r(t), s^{\prime}\}$;\\
Randomly select $M_b$ transitions from the replay buffer and input them to the actor and critic; \\
Update the evaluation networks' parameters for the actor and critic according to policy gradient and loss function;\\
Update the target networks' parameters with (\ref{soft_ac}) and (\ref{soft_cri}).\\}
$r_{ep} = r_{ep} + r(t)$.
 }
 $s_0=s(t)$\\
 }
\caption{The DDPG-based solution
\label{DDPG_Algorithm}}
\end{algorithm}

\section{Simulation Results}
\label{sec:Simulation_Res}

To demonstrate the performance of the DDPG-based resource management scheme, we simulate vehicle mobility with planung transport verkehr (PTV) Vissim first, and then use the obtained traffic data and the randomly generated computing tasks to train and test the DDPG-based resource management model. Assume the vehicle's transmit power is $1\,$Watt. The channel gains of uplink transmissions from a vehicle to the MeNB and from a vehicle to UAV $j$ are described as $L(d_{m,i}) = -30-35\rm{log}_{10}(d_{m,i})$ and $L(d_{j,i}) = -40-35\rm{log}_{10}(d_{j,i})$ \cite{Ye2018Dynamic}, respectively, where, $d_{m,i}$ (or $d_{j,i}$) is the vehicle-MeNB (or vehicle-UAV) distance. Considering the heterogeneous vehicular applications, we assume vehicle $i \in \mathcal{N}(t)$ periodically generates different computing tasks with $c_i^{co}(t) \in [50,100]\,$MHz, $c_i^{da}(t) \in [0.5, 1]\,$kbits, and $c_i^{de}(t) \in [10, 50]\,$ms. Unless otherwise stated, all other parameters used in the training and inferring stages are set in Table \ref{table:parameters}.

\begin{table}[htbp]
\setlength{\belowcaptionskip}{5pt}
 \caption{\label{table:parameters} Parameters set for the training stage}
 \centering
 \begin{tabular}{p{5.3cm}||p{2cm}}
  \toprule
  \textbf{Parameter} & \textbf{Value} \\
  \midrule
  Height of the MeNB & $50\,$m \\
  Altitude of the UAV & $40\,$m \\
  Flying speed of the UAV & $10\,$m/s \\
  Available spectrum resources for the MeNB/UAV & $10/2\,$MHz \\
  Computational capabilities at the MeNB/UAV & $250/30\,$GHz \\
  Caching resources at the MeNB/UAV & $50/6\,$kbits \\
  Communication range of the MeNB/ UAV & $600/100\,$m \\
  Background noise power & $-104\,$dBm \\
  Replay buffer size & $10000$ \\
  Size of each mini-batch of transitions & $32$ \\
  Discount factor on reward & $0.9$ \\
  Learning rate of the actor/critic & $0.0005/0.005$ \\
  $\kappa_a$/$\kappa_c$ & $0.05$ \\
  \bottomrule
 \end{tabular}
\end{table}

\begin{figure}[htbp]
\centering
\includegraphics[scale=0.29]{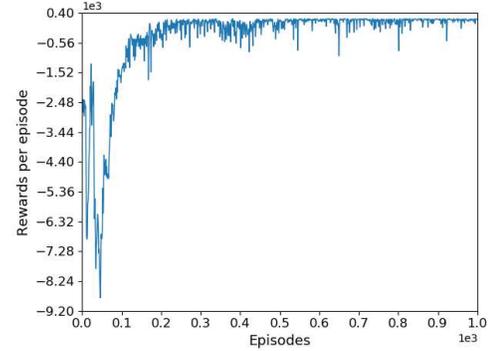}
\caption{Total rewards achieved per episode during the training stage.}
\label{fig:converge}
\end{figure}

Fig. \ref{fig:converge} demonstrates the convergence performance of the presented DDPG-based resource management scheme. Since the parameters of the DDPG model are initialized according to state $s_0$ at the beginning of the training stage and the corresponding policy is not optimal, total rewards achieved in the first $200$ episodes are relatively small and fluctuate dramatically. With the training going, the policy tends to be optimal. Thus, high total rewards are achieved and the reward fluctuation is getting smaller and smaller in the last $800$ episodes. As $1100$ steps are included in each episode and the two reward elements $r_i^{de}$ and $r_i^{ca}$ are clipped to $0.1$, the highest achieved total rewards during the training stage is $220$, as shown in Fig. \ref{fig:converge}. By clipping the two reward elements to $0.1$, the highest rewards achieved by each vehicle are limited to $0.2$. Thus, high rewards achieved by vehicle $i$ from excessive allocated resources can be avoided, therefore guiding the agent to learn the optimal policy fast and improving the converge performance of the DDPG-based solution. 


For the DDPG-based solution, vehicle association and resource management decisions made by the optimal policy should maximize the long-term rewards. As positive rewards are achieved by vehicles only when its offloaded task's delay or QoS requirements are satisfied by the allocated resources, high episode rewards indicate more offloaded tasks are with satisfied delay/QoS requirements. Hence, we use two evaluation criterion, delay/QoS satisfaction ratios\footnote{The delay/QoS satisfaction ratios are the ratios of the tasks completed by the MEC servers with satisfied delay/QoS requirements over the total number of offloaded tasks \cite{peng2020deep}.}, to measure the performance of the DDPG-based resource management scheme. Also, we compare the DDPG-based scheme with the random scheme, which rapidly and randomly decides the association patterns for and allocates resources to vehicles under the coverage area of the MeNB. 

\begin{figure}[htbp]
\centering
\subfigure[Vs. spectrum resources]{
\label{subfig:vs_spe}
\includegraphics[scale=0.28]{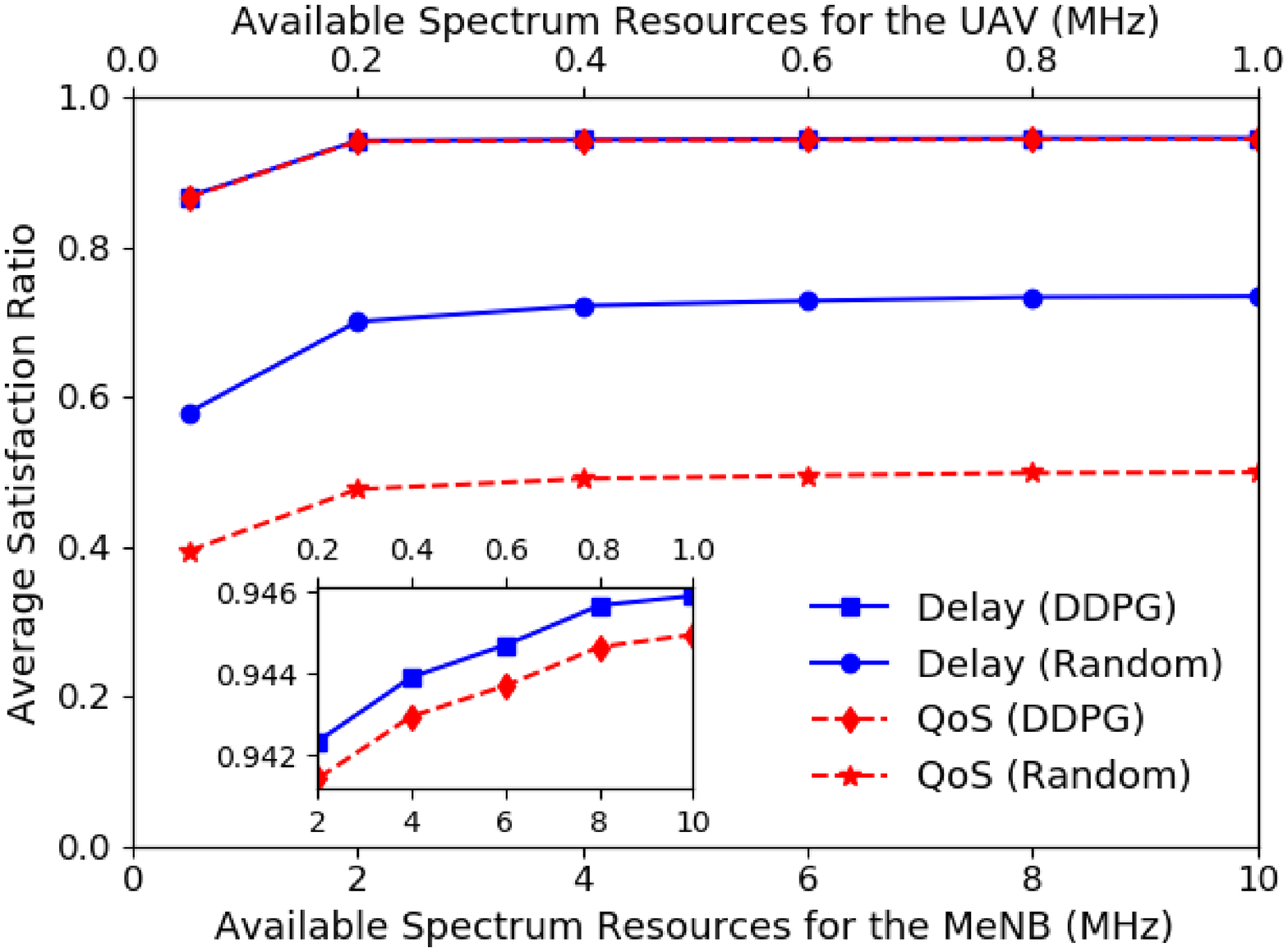}}
\subfigure[Vs. computation capabilities]{
\label{subfig:vs_comp}
\includegraphics[scale=0.28]{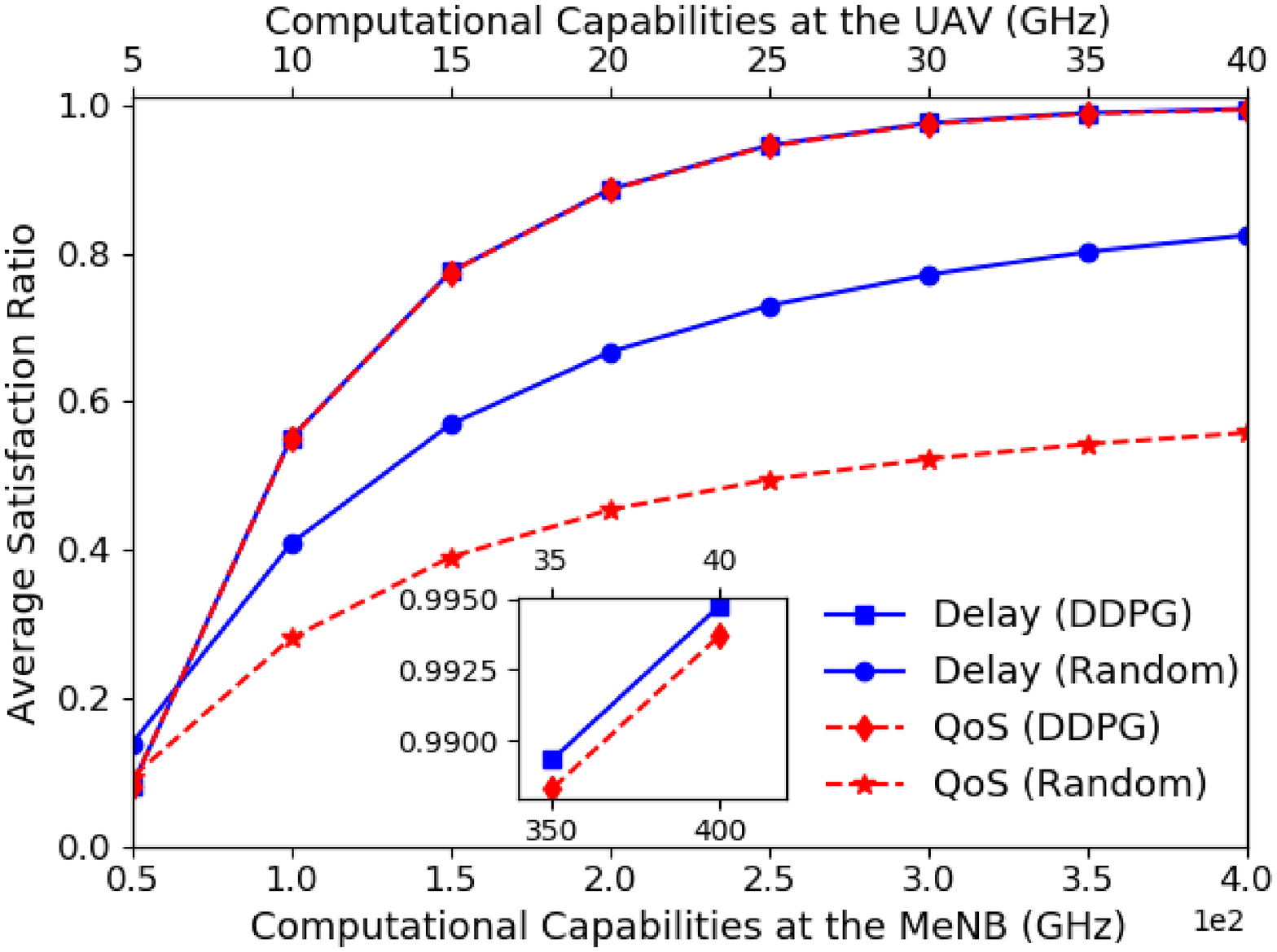}}
\subfigure[Vs. caching resources]{
\label{subfig:vs_stor}
\includegraphics[scale=0.28]{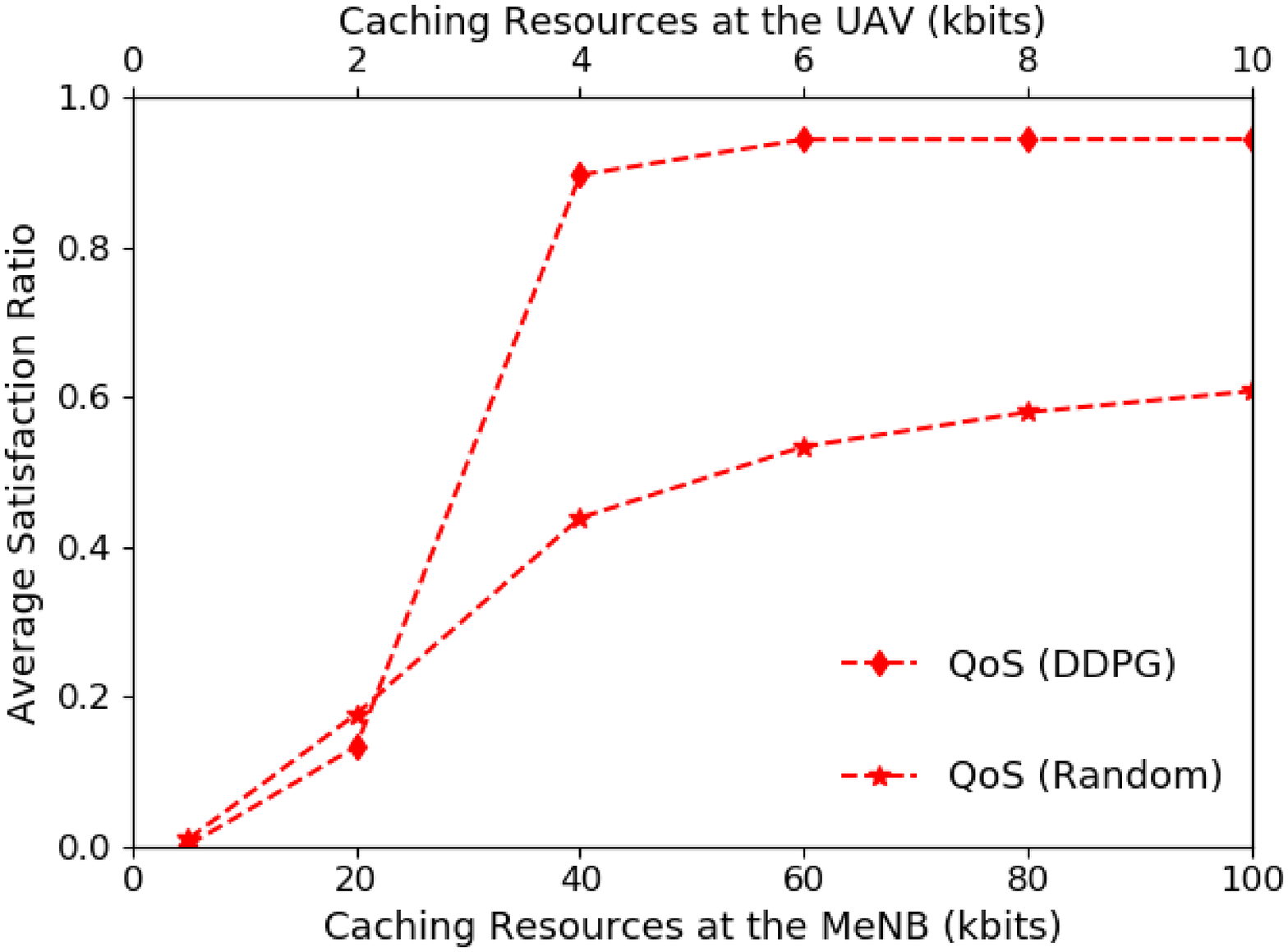}}
\caption{Average delay/QoS satisfaction ratios over 10000 tests.}
\label{fig:Satis_ra}
\end{figure}

The three subfigures in Fig. \ref{fig:Satis_ra} show the average delay/QoS satisfaction ratios over $10,000$ times of tests versus the amounts of available spectrum, computing, and caching resources for the MeNB and UAVs. With more available resources for the MeNB and UAVs, resources allocated to each task offloading request are increased, and therefore resulting in more tasks completed by the MEC servers with satisfied delay/QoS requirements. For each offloaded task, the satisfied QoS requirement indicates the delay requirement is satisfied and at least $c_i^{da}$ caching resources are allocated to that task. Thus, under the same test setting, the average delay satisfaction ratio is always higher than the QoS one, as shown in Fig. \ref{fig:Satis_ra}. Moreover, we can see that the gap between the delay and the QoS satisfaction ratios of the DDPG-based scheme is far less than that of the random scheme because the DDPG-based scheme jointly allocates the multi-dimensional resources to satisfy the offloaded tasks' QoS requirements.


\section{Conclusion}
\label{sec:Conclu}

This paper has investigated the multi-dimensional resource management problem in the MEC/UAV-assisted vehicular network. Particularly, to support as many offloaded computing tasks with satisfied delay and QoS requirements as possible, an optimization resource management problem has been formulated for the controller installed at the MeNB. As the formulated problem is non-convex and with high computational complexity, instead of leveraging traditional optimization methods, we have transformed the original problem with RL and designed a DDPG-based solution. Simulation results have demonstrated that the DDPG-based algorithm can converge within $200$ episodes during the training stage and higher delay/QoS satisfaction ratios can be achieved by the DDPG-based scheme than the random scheme. 

\bibliographystyle{IEEEtran}
\bibliography{Confe_SADDPG}

\end{document}